\documentclass[a4paper,9pt]{sig-alternate}

\newcommand{\mbf}{\mathbf}
\newcommand{\wtd}{\widetilde}

\usepackage[ruled,linesnumbered]{algorithm2e}
\usepackage{url}
\usepackage{multirow}
\usepackage{graphicx}
\usepackage{amssymb}
\usepackage{amsmath}
\usepackage{amsfonts}
\usepackage{color}
\usepackage{epstopdf}
\usepackage{verbatim} 
\usepackage[noadjust]{cite}
\usepackage{threeparttable}

\usepackage{soul}
\usepackage{verbatim, url}

\newcounter{theorem}
\newcounter{def}
\newcounter{mylea}
\newcounter{corollary}
\usepackage{fancyhdr}

\newtheorem{lemma}[mylea]{Lemma}

\begin{document}
\title{MATEX: A Distributed Framework for\\ Transient Simulation of Power Distribution Networks}
\author{Hao~Zhuang$^\S$, 
        Shih-Hung~Weng$^\dagger$, 
        Jeng-Hau Lin$^\S$, 
        Chung-Kuan~Cheng$^\S$\\\\
        $^\S$Computer Science \& Engineering Department, University of California, San Diego, CA, 92093\\
        $^\dagger$Facebook Inc., Menlo Park, CA, 94025\\
         zhuangh@ucsd.edu,~shweng@fb.com,~jel252@ucsd.edu,~ckcheng@ucsd.edu
        }
        \vspace{-1in}
\maketitle
\begin{abstract}

We proposed \emph{MATEX}, a distributed framework for
transient simulation of power distribution networks (PDNs).
MATEX utilizes \underline{mat}rix \underline{ex}ponential kernel
with Krylov subspace approximations to solve differential equations of linear circuit.
First, the whole simulation task is divided into subtasks 
based on decompositions of current sources, in order to reduce the computational overheads. 
Then these subtasks are distributed to different computing nodes and 
processed in parallel.
Within each node, after the  matrix factorization at the beginning of simulation, the adaptive time stepping solver
is performed without extra matrix re-factorizations.
MATEX overcomes the stiffness hinder of previous 
matrix exponential-based circuit simulator 
by \emph{rational} Krylov subspace method, which leads to larger step sizes with  smaller dimensions of Krylov subspace bases and highly accelerates the whole computation.
MATEX outperforms both traditional fixed and adaptive time stepping methods,
e.g., achieving around 13X over the trapezoidal framework with fixed time step for the IBM power grid benchmarks.

\end{abstract}

{
\scriptsize
\section*{Categories and Subject Descriptors} 
B.7.2 [Integrated Circuits]: Design Aids, Simulation. 
J.6 [Computer-aided engineering]: Computer-aided design (CAD).

\section*{General Terms} 
Algorithms,  Design, Theory, Verification, Performance

\section*{Keywords}
 Circuit Simulation, 
 Power Distribution Networks, 
 Power Grid, 
 Transient Simulation,
 Matrix Exponential, 
 Krylov Subspace, 
 Distributed Computing, Parallel Processing.
}

\section{Introduction}
\label{sec:intro}

Modern VLSI design verification relies heavily 
on the analysis of power distribution network (PDN)
to estimate power supply noises. 
PDN is often modeled as a large-scale linear circuit with voltage
supplies and time-varying current sources \cite{Zhao02, Li12_Tau}. 
Such circuit is extremely large, which makes 
the corresponding transient simulation very time-consuming.
Therefore, scalable and theoretically elegant algorithms
for the transient simulation of linear circuits have been always favored.
Nowadays, the emerging multi-core, many-core platforms bring powerful 
computing resource and opportunities for parallel computing. 
Even more, 
cloud computing techniques \cite{cloud} drive 
distributed systems scaling to thousands of computing nodes
\cite{mesos}, etc.
Such distributed systems will be also promising 
computing resources in EDA industry.
However, building scalable and efficient distributed algorithmic framework for 
transient linear circuit simulation framework is still a challenge to leverage these powerful computing tools.
Previous works 
\cite{Li11, Weng12_ICCAD, Weng12_TCAD, Wang13}
have been made in order to improve circuit simulation by novel algorithms, 
parallel processing and distributed computing.

Traditional numerical methods solve differential algebra equations (DAEs) explicitly, e.g., forward Euler, or implicitly, 
e.g., backward Euler (BE), trapezoidal (TR) method, which are based on low order polynomial approximations.
Due to the stiffness of systems, which comes from a wide range of time constants of a circuit,
the explicit methods require small time step sizes to ensure the stability.
In contrast, implicit methods can deal with this problem because of their larger stability regions. 
However, at each time step, these methods have to solve a linear system, which is sparse and often ill-conditioned.
Due to the requirement of a robust solution,
compared to iterative matrix solvers \cite{Saad03},
direct matrix solvers \cite{Davis06} 
are often favored for VLSI circuit simulation, and thus adopted by
state-of-the-art power grid (PG) solvers in TAU PG simulation contest \cite{Yu12,Yang12,Xiong12}. 
During transient simulation, these solvers require only one matrix factorization (LU or Cholesky factorization) at the beginning.  
Then, the following transient computation, at each fixed time step, needs only a pair of forward and backward substitutions, which achieves better efficiency over adaptive stepping methods by reusing the factorized matrix \cite{Xiong12, Yu12, Li12_Tau}. 

Beyond traditional methods, 
a new class of methods called exponential
time differencing (ETD) has been embraced by MEXP
\cite{Weng12_TCAD}.
The major complexity of ETD is caused by matrix exponential computations. 
MEXP utilizes standard Krylov subspace method based on \cite{Saad92} to approximate matrix exponential and vector product. MEXP can solve the DAEs with high polynomial approximations \cite{Weng12_TCAD, Saad92} than traditional ones.
Another merit of using MEXP-like SPICE simulation for linear circuit is 
the adaptive time stepping, which can proceed without
re-factorizing matrices on-the-fly, while the traditional counterparts cannot avoid such time-consuming process during the adaptive time marching.
Nevertheless, when simulating stiff circuits,
utilizing standard Krylov subspace method requires large dimension of basis 
in order to preserve the accuracy of MEXP approximation. 
It may pose memory bottleneck and degrade the adaptive stepping performance of MEXP.

In this paper,
we propose a distributed algorithmic framework for PDN transient simulation, called as \emph{MATEX}, which inherits the matrix exponential kernel. 
First, the PDN's input sources are partitioned into groups based on their similarity. They are assigned to different computing nodes to run the corresponding PDN transient simulations. Then, the results among nodes are summed up, according to the well-known superposition property of linear system.
This partition reduces the chances of generating Krylov subspaces and enlarges the time periods of reusing them during the transient simulation at each node, which brings huge computational advantage. 
In addition, we also highly accelerate the circuit solver by adopting \emph{inverted} and \emph{rational} Krylov subspace methods for the computation of matrix exponential and vector product. We find the rational Krylov subspace method is the most efficient one, which helps MATEX leverage its flexible adaptive time stepping by reusing factorized matrix at the beginning of transient simulation. 
In IBM power grid simulation benchmarks, 
our framework gains around \textbf{13X} speedup on average
in transient computing part after its matrix factorization,
compared to the commonly adopted TR method with fixed time step.
The overall speedup is \textbf{7X}.

\emph{Paper Organization}.
Section \ref{sec:method} introduces the background of
linear circuit simulation and matrix exponential formulations.
Section \ref{sec:matex} presents overall framework of MATEX.  
Section \ref{sec:exp_results} shows numerical results 
and Section \ref{sec:conclusion} concludes this paper.

\section{Preliminaries}
\label{sec:method}

\subsection{Transient  Simulation of Linear Circuits}
Transient linear circuit simulation is the foundation of PDN simulation. It is formulated as DAEs via modified nodal analysis (MNA),
\begin{eqnarray}
    \label{eqn:dae}
\footnotesize 
    \mbf{C}\dot{\mbf{x}}(t) = - \mbf{G}\mbf{x}(t) + \mbf{B}\mbf{u}(t),
\end{eqnarray}
where  $\mbf{C}$ is the  matrix resulting from capacitive and inductive
elements.
$\mbf{G}$ is the  conductive matrix, 
and $\mbf{B}$ is the input selector matrix. 
$\mbf{x}(t)$ is the vector of time-varying node voltages
and branch currents. 
$\mbf{u}(t)$  is the vector of  
supply voltage and current sources. In PDN, such current sources
are often characterized as pulse inputs \cite{Nassif08_Power, Li12_Tau}.
To solve Eq. (\ref{eqn:dae}) numerically,
it is, commonly, 
discretized with time step $h$ and transformed
to a linear algebraic system. 
Given an initial condition $\mbf x(0)$ from DC analysis, 
or previous time step $\mbf x(t)$,
For a time step $h$,
$\mbf x(t+h)$ can be obtained by traditional \emph{low order approximation} methods, e.g., TR, 
which is an implicit second-order method, and probably most commonly used
strategy for large scale circuit simulation.
\begin{eqnarray}
    \label{eqn:trap}
    (\frac{\mbf{C}}{h} + 
    \frac{\mbf G}{2}) {\mbf{x}}(t+h) 
    = (\frac{\mbf C}{h}- \frac{\mbf G}{2} )
    \mbf{x}(t) + \mbf{B} \frac{ \left ( \mbf{u}(t) + \mbf{u}(t+h) \right)} {2}
\end{eqnarray}
 
Besides, TR with fixed time step $h$ is an efficient framework and adopted by the top PG solvers in 2012 TAU PG simulation contest \cite{Yu12, Yang12, Xiong12, Li12_Tau}.

\subsection{Exponential Time Differencing Method}
The solution of Eq. (\ref{eqn:dae}) can be obtained analytically \cite{Leon75}. 
For simple illustration, we convert Eq. (\ref{eqn:dae}) into 
\begin{eqnarray}
\footnotesize
\label{eqn:ode}
    \dot{\mbf{x}}(t) = \mbf{A}\mbf{x}(t) + \mbf{b}(t),
\end{eqnarray}
when $\mbf{C}$ is not singular, 
$\mbf{A} = - \mbf{C}^{-1}\mbf{G}$ 
and $\mbf{b}(t) = \mbf{C}^{-1}\mbf{B}\mbf{u}(t)$.
Given the solution at time 
$t$ and a time step $h$, the solution at
$t + h$ is
\begin{eqnarray} 
\footnotesize
    \label{eqn:discrete_sol}
    \mbf{x}(t+h) = e^{h\mbf{A}}\mbf{x}(t) + 
    \int^{h}_{0}e^{(h-\tau)\mbf{A}}\mbf{b}(t+\tau)d\tau.
\end{eqnarray}

Assuming that the input $\mbf{u}(t)$ is piecewise linear (PWL), e.g.
$\mbf u(t)$ is linear within every time step, 
we can integrate the last term of  
Eq. (\ref{eqn:discrete_sol}), analytically, 
turning the solution with matrix exponential operator:
\begin{eqnarray} 
\label{eqn:pwl_exact_sol}
    \mbf{x}(t+h)
    &=& 
    e^{h\mbf{A}}(\mbf{x}(t) + \mbf A^{-1}\mbf b(t) + \mbf A^{-2} \frac{\mbf b(t+h)- \mbf b(t)}{h})
    \nonumber \\
    & & 
    -(\mbf A^{-1}\mbf b(t+h) + \mbf A^{-2} \frac{\mbf b(t+h)- \mbf b(t)}{h})
\end{eqnarray}
For the time step choice, \emph{input transition spots (TS)} refer to the time points where slopes of input sources vector changes. Therefore, for Eq. (\ref{eqn:pwl_exact_sol}), the maximum time step starting from $t$ is $(t_s-t)$,
where $t_s$ is the smallest one in $TS$ 
larger than $t$.


In Eq. (\ref{eqn:pwl_exact_sol}), $\mbf A$ in 
$e^{\mbf A}\mbf v$ is usually above millions, making the direct computation infeasible. 

\subsection{Matrix Exponential Computation by Standard Krylov Subspace Method}

The complexity of $e^{\mbf A}\mbf v$ can be reduced 
using Krylov subspace method and still maintained  
in a high order polynomial approximation \cite{Saad92}, 
which has been deployed by MEXP \cite{Weng12_TCAD}.
In this paper, we call the Krylov subspace utilized in MEXP as \emph{standard Krylov subspace}, due to its
straightforward usage of $\mbf A$ when generating basis through 
Arnoldi process in Alg. \ref{algo:arnoldi}. 
First,
we reformulate Eq. (\ref{eqn:pwl_exact_sol}) into
\begin{eqnarray}
\label{eqn:new_exact}
    \mbf{x}(t+h)
    =
    e^{h\mbf A}(\mbf x(t) + \mbf F(t,h) ) - \mbf P (t,h) 
\end{eqnarray}
where 
$\mbf F(t,h) =  \mbf A^{-1}\mbf b(t) + \mbf A^{-2} \frac{\mbf b(t+h)- \mbf b(t)}{h}$
and  $ \mbf P(t,h) = (\mbf A^{-1}\mbf b(t+h) + \mbf A^{-2} \frac{\mbf b(t+h)- \mbf b(t)}{h})$.
The standard Krylov subspace 
${\rm K_m}(\mbf A, v) := \text{span}
{\{ \mbf v, \mbf A\mbf v,
\cdots, \mbf A^{m-1} \mbf v}\}$
obtained by Arnoldi process
has the relation
$\mbf{A}\mbf{V_m} = 
\mbf V_m \mbf H_m
+ h_{m+1,m} 
\mbf v_{m+1}\mbf e_m^\mathsf{T} 
$, where $h_{m+1,m}$ is the $(m+1,m)$ entry of Hessenberg matrix
$\mbf H_m $, and $e_m$ is the $m$-th unit vector.
The matrix exponential and vector product is computed via 
$ e^{h\mbf A} \mbf v \approx \lVert \mbf v \rVert \mbf V_m e^{h\mbf {H}_m} \mbf e_1$. 
The $\mbf H_m$ is usually much smaller compared to $\mbf A$.
The posterior error term is 
\begin{eqnarray}
\label{eqn:std_error}
\lVert r_m(h) \rVert = \lVert \mbf v \rVert \left|  h_{m+1,m}  \mbf v_{m+1} 
\mbf e^{\mathbf T}_m e^{h\mbf{H}_m} \mbf e_1 \right |
\end{eqnarray}
To generate $\mbf x(t+h)$ by Alg. \ref{algo:arnoldi}, 
we use $[\mbf L, ~\mbf U] = \text{LU\_Decompose}(\mbf X_1)$, where, for
standard Krylov subspace, $\mbf X_1 = \mbf C$, and
$\mbf X_2 = \mbf G$ as inputs. 
The error budget $\epsilon$ and Eq. (\ref{eqn:std_error}) are used 
to determine the convergence condition in current time step $h$ with an order $j$ of Krylov subspace dimension for $e^{\mbf A}\mbf v$ approximation (from line \ref{alg1:conv1} to line \ref{alg1:conv2}). 
\begin{algorithm}
\label{algo:arnoldi}
\caption{MATEX\_Arnoldi}
\KwIn{ $\mbf L, \mbf U, \mbf X_2, h ,t, \mbf x(t),  \epsilon, \mbf P (t,h),\mbf F (t,h)$  }
\KwOut{ $\mbf x(t+h), \mbf V_m, \mbf H_m, \mbf v$}
    {   
	$\mbf v = \mbf x(t) + \mbf F(t,h) , \mbf v_1 = \frac{\mbf v }{\lVert \mbf v \rVert}$\;
	\For  {$j=1:m$}
	{
	    $\mbf w = \mbf U \backslash  (\mbf L \backslash (\mbf X_2 \mbf v_{j}))$
	    {\tcc*[r]{a pair of forward and backward  substitutions}}
	    \For {$i = 1:j$}
	    {
		$h_{i,j} = \mbf w^T\mbf v_{i}$\;
		$\mbf w = \mbf w - h_{i,j} \mbf v_{i}$\;
	    }
	    $h_{j+1,j} = \lVert \mbf w \rVert $\;
	    $\mbf v_{j+1} = \frac{\mbf w }{h_{j+1,j}} $\;
	    \If {$ || \mbf r_j(h)|| < \epsilon $\label{alg1:conv1}} 
	    {
		$m = j$; break\; 
	    }\label{alg1:conv2}
	}
	$\mbf x(t+h) = \lVert \mbf v \rVert \mbf {V_m} e^{h \mbf H_m} \mbf e_1 - \mbf P(t,h)$\;
	\label{alg1:h}
    }
\end{algorithm}

\subsection{Discussions of MEXP}
\label{sec:discuss}
The input term $\mbf b$ embedded in Eq. (\ref{eqn:discrete_sol}) serves a double-edged sword in MEXP.
First, the flexible time stepping can choose any time spot until the next input transition spot $t_s$, as long as the 
approximation of $e^{\mbf A}\mbf v$
is accurate enough. The Krylov subspace can be reused 
when $t+h \in [t,t_s]$, only by scaling $\mbf H_m$ with $h$ 
in $\mbf x(t+h) = \lVert \mbf v \rVert \mbf {V_m} e^{h \mbf H_m} \mbf e_1 - \mbf P(t,h)$. 
This is an important feature that even doing the adaptive time stepping, we can still use the last Krylov subspaces.

However,
the region before the next transition $t_s$
may be shortened
when there are a lot of independent input sources injected into the linear system. It leads to more chances of generating new Krylov subspace.
This issue is addressed in Sec. \ref{sec:frame_motivation} and
Sec. \ref{sec:all_frame}.

The standard Krylov subspace may not be 
computationally efficient  
when simulating stiff circuits
based on MEXP\cite{Weng12_TCAD, Weng12_ICCAD}.
For the accuracy of approximation of $e^{\mbf A}\mbf v$, large dimension of Krylov subspace basis is required, which not only brings the computational complexity but also consumes huge memory.
Besides, for a circuit with singular $\mbf C$, during the generation of standard Krylov subspace, a
\emph{regularization} process is required to convert such  $\mbf C$ into non-singular one, which is time-consuming for large scale circuits. 
These two problems are solved in Sec. \ref{sec:advanced}.


\section{MATEX Framework}
\label{sec:matex}
\begin{figure}[ht] 
\scriptsize
    \centering
    \includegraphics[width=3.4in ]{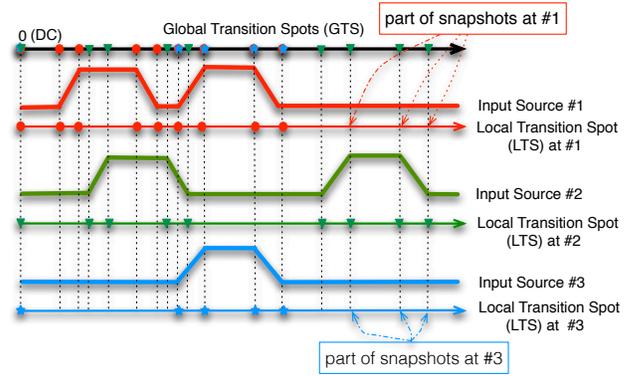}
    \caption{
	{
	    \scriptsize
	{Illustration of input transitions. 
	GTS: Global Transition Spots; LTS: Local Transition Spots;
	Snapshots: the crossing positions by dash lines and LTS \#$k$ without solid points. 
	}
	}
    }
    \label{fig:input}
\end{figure}
\begin{figure}[ht]
\scriptsize
    \centering
    \includegraphics[ width=1.9in]{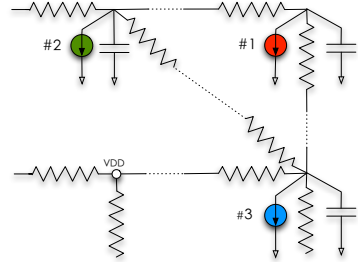}
    \caption{Part of a PDN model with input sources from Fig. \ref{fig:input}}
	\label{fig:rc_ckt}
\end{figure}
\begin{figure}[ht]
\scriptsize
    \centering
    \includegraphics[ width=3.4in ]{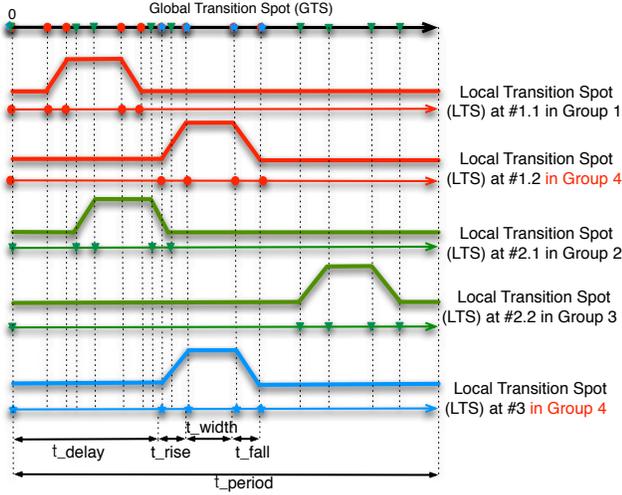}
    \caption{
	\scriptsize
	{Grouping of ``bump'' shape transitions for sub-task simulation.
	The matrix exponential-based method can utilize  adaptive stepping in each LTS and reuse Krylov subspace generated at the most recent solid point. 
	However, traditional methods (TR, BE, etc) still need to do time marching step by step, either by pairs of  forward and backward substitutions to proceed with fixed time step, or re-factorizing matrix and solving linear system when using adaptive stepping strategy. 
	    (Pulse input information:
	        t\_delay: initial delay time;
	        t\_rise: rise time;
	        t\_width: width of pulse-wise; 
	        t\_fall: fall time;
	        t\_period: period).
	}
    }
	\label{fig:ckt_ts_decomp}
\end{figure}

\vspace{-0.03in}
\subsection{Motivation}
\label{sec:frame_motivation}

Matrix exponential kernel with Kyrlov subspace method 
can solve Eq. (\ref{eqn:dae}) with larger time steps than 
 lower order approximation methods. 
Our motivation is to leverage this advantage to reduce the number of time step and accelerate the transient simulation.   
However, there are usually many input currents in PDNs, which
narrow the regions for the  time stepping of matrix exponential-based method. We want to utilize the well-known
superposition property of linear system and distributed computing model to tackle this challenge.
To illustrate our framework briefly, 
we first define three terms:

\textbf{Definition}: \emph{Local Transition Spot} ($LTS$) is the set of $TS$ at an input source to the PDN.

\textbf{Definition}: \emph{{Global Transition Spot}} ($GTS$) is the union of $LTS$ among all the input sources to the PDN. 

\textbf{Definition}: \emph{{Snapshot}} denotes a set $GTS \setminus LTS$ at an input source.

If we simulate the PDN with respective to all the inputs,
$GTS$ are the places where generations of Krylov subspace 
cannot be avoided. 
For example, there are three input sources in a PDN (Fig. \ref{fig:rc_ckt}). The input waveforms are shown in Fig. \ref{fig:input}.
Then, the first line is that $GTS$, which is contributed by all $LTS$ from input sources \#1, \#2 and \#3.

However, we can partition the task to sub-tasks by simulating each
input sources individually. Then, each
sub-task only needs to generate Krylov subspaces
based on its own $LTS$ and keep track of $Snapshot$
for the later usage of summation via superposition. 
In addition, the points in \emph{Snapshot} between two points $l_1, l_2 \in LTS$ ($l_1 < l_2$), can reuse the Krylov subspace generated at $l_1$, which is mentioned in Sec. \ref{sec:discuss}.
For each node, the chances of Krylov subspaces generations are reduced and the time
periods of reusing these subspaces are enlarged locally, which bring
huge computational benefit when processing these subtasks in parallel.
 

Above, we divide the simulation task by input sources.
 We can, more aggressively, decompose the  task according to the ``bump'' shapes within such input pulse  sources. 
We group the ones which have the same (t\_delay, t\_rise, t\_fall, t\_width) into one set, 
which is shown in Fig. \ref{fig:ckt_ts_decomp}. 
There are 4 groups in Fig. \ref{fig:ckt_ts_decomp},
Group 1 contains LTS\#1.1, Group 2 contains LTS\#2.1, Group 3 contains LTS \#2.2, 
and Group 4 contains LTS \#1.2 and \#3.

\vspace{-0.03in}
\subsection{MATEX Framework}
\label{sec:all_frame}
\begin{figure}[h]
\centering
\includegraphics[ width=3.4in ]{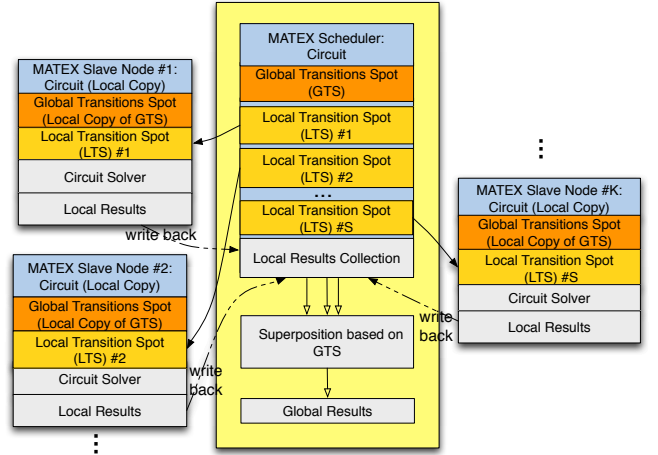}
\caption{The flow of MATEX framework}
\label{fig:matex}
\end{figure}

Our proposed framework MATEX is shown in Fig. \ref{fig:matex}. After pre-computing $GTS$ and decomposing
$LTS$ based on ``bump'' shape (Fig. \ref{fig:ckt_ts_decomp}), we group them and form 
$LTS$ $\#1 \sim \#K$ (\emph{Note: there are alternative decomposition strategies. It is also easy to extend the work 
to deal with different input waveforms. We try to keep this part as simple as possible to emphasize our framework}).  

MATEX scheduler sends out $GTS$ and $LTS$ to different MATEX slave node.
Then the simulations are  processed in parallel.
There are no communications among nodes before the ``write back''.
Within each slave node, ``circuit solver'' (Alg. \ref{algo:ckt_solver}) 
computes transient response with varied time steps. 
Solutions are obtained without re-factorizing matrix during the transient computing. 
After finishing all simulations from slave nodes, they writes back the results
and informs the MATEX scheduler.

\vspace{-0.03in}
\begin{algorithm}
\label{algo:ckt_solver}
\caption{MATEX Circuit Solver Algorithm}
\KwIn{ $LTS$ \#$k$,  $GTS$, $\mbf X_1$, $\mbf X_2$, and $\mbf P_k$, $\mbf F_k$, which contain the corresponding $\mbf b$ for node $k$. 
}
\KwOut{Local solution $\mbf x$ along $GTS$ in node 
$k \in [1,\cdots,S]$, where $S$ is the number of nodes}
    {   
	$t=T_{start}$\;
	$\mbf x(t) = \text{Local\_Initial\_Solution}$\;
	$[\mbf L,\mbf U] = \text{LU\_Decompose}( \mbf X_1)$\;
	\While { $t \leq T_{end}$}
	{
	Compute maximum allowed step size $h$ based on $GTS$\;
	\If{$t \in $  LTS \#$k$}
	{
	\tcc{Generate the Krylov subspace for the time point of $LTS$ and compute $\mbf x$}
	$[\mbf x(t+h), \mbf V_m, \mbf H_m,\mbf v]
	= \text{MATEX\_Arnoldi}$$(\mbf L,\mbf U,\mbf X_2, h, t, \mbf x(t), \epsilon, \mbf P_k(t,h), \mbf F_k(t,h))$;
	\label{algo2:arnoldi}
	$a_{lts} = t$; 
	}
	\Else{
	\tcc{Compute the $\mbf x$ at $Snapshot$ by reusing the latest Krylov subspace}
	    {
	    $h_{a} = t + h - a_{lts}$\;
	    $\mbf x(t+h) = \lVert \mbf v \rVert \mbf {V_m} e^{h_{a} \mbf H_m} \mbf e_1 - \mbf P_k(t,h)$\;
	    \label{algo2:h}
	    }
	}
	$t= t+h$\;
	}
    }
\end{algorithm}


\subsection{Circuit Solver Accelerations}
\label{sec:advanced}

As mention in Sec. \ref{sec:discuss}, standard Krylov subspace 
approximation in MEXP \cite{Weng12_TCAD} is not computationally
efficient for stiff circuit.
The reason is that Hessenberg matrix $\mbf H_{m}$ of standard Krylov subspace tends to approximate the large magnitude eigenvalues of $\mbf A$ \cite{Van06}.
Due to the exponential decay of higher order terms in Taylor's expansion, such components are not the crux 
of circuit system's behavior 
\cite{Botchev12, Van06}.
Dealing with stiff circuit, therefore, needs to gather more vectors into subspace basis 
and increase the size of $\mbf H_{m}$
to fetch more useful components, which results to both
memory overhead and computational complexity into 
Krylov subspace generations during time stepping.
Direct deploying MEXP into MATEX's \emph{Circuit Solver} 
is not efficient to leverage the benefits of local flexible and larger time stepping.
In the following subsections, we adopt the idea from \emph{spectral transformation}
\cite{Botchev12, Van06}
to effectively capture small magnitude eigenvalues in $\mbf A$,
leading to a fast yet accurate \emph{Circuit Solver} for MATEX.

\subsubsection{Matrix Exponential and Vector Computation by Inverted Krylov Subspace (I-MATEX)}
Instead of $\mbf A$, we use $\mbf A^{-1}$ (or -$\mbf G^{-1} \mbf C$) as our target matrix to form
${\rm K_m}(\mbf A^{-1} , v) := \text{span}
{\{ \mbf v, \mbf A^{-1}\mbf v, \cdots, \mbf A^{-(m-1)} \mbf v}\}
$.
Intuitively, by inverting $\mbf A$, the small magnitude eigenvalues become the large ones
of $\mbf A^{-1}$.
The resulting $\mbf H'_{m}$ is likely to capture these
eigenvalues first. 
Based on Arnoldi algorithm, 
the inverted Krylov subspace has the relation 
$ \mbf{A}^{-1}\mbf{V_m} =  
     \mbf V_m \mbf H'_m + 
     h'_{m+1,m} \mbf v_{m+1}\mbf e_m^\mathsf{T} 
     $.
The matrix exponential $e^{\mbf A} \mbf v$ is calculated as
$ \lVert \mbf v \rVert \mbf V_m e^{h\mbf {H'}^{-1}_m} \mbf e_1 $.
To put this method into Alg. 1, just by modifying the input variables, 
$\mbf X_1 = \mbf G$ for the LU decomposition,
and $\mbf X_2 = \mbf C$.  
In the line \ref{alg1:h} of Alg. \ref{algo:arnoldi}, $\mbf H_m = \mbf {H'}^{-1}_m$.
The posterior error approximation is
\begin{eqnarray}
\label{eq:err_inverted_krylov}
  \lVert \mbf r_m(h) \rVert 
  =
  \lVert \mbf v  \rVert
  \left |  \mbf A  h'_{m+1,m} \mbf v_{m+1} 
  \mbf e^T_m {\mbf H'}^{-1}_{m}
  e^{h {\mbf H'}^{-1}_{m}}
  \mbf e_1 \right |
\end{eqnarray}
which is derived from residual-based error approximation in
\cite{Botchev12}.

\subsubsection{Matrix Exponential and Vector Computation 
by Rational Krylov Subspace (R-MATEX)}
The shift-and-invert Krylov subspace basis 
\cite{Van06} 
is designed
to confine the spectrum of
$\mbf A$.
Then, we generate Krylov subspace via
$    \mbf {K_m}((\mbf I- \gamma \mbf {A} )^{-1}, \mbf v)   =   
	    span\{ \mbf v, (\mbf I- \gamma\mbf { A} )^{-1}\mbf v,
    \cdots,
      (  \mbf I -  \gamma \mbf{A})^{-(m-1)} \mbf v \},
      $
where $\gamma$ is a predefined parameter.
With this shift, all the eigenvalues' magnitudes are larger than one. 
Then the invert limits the magnitudes smaller than one.
According to \cite{Botchev12, Van06},
the shift-and-invert basis for matrix exponential-based transient simulation
is not very sensitive to $\gamma$, once it is set to around the order near time steps 
used in transient simulation.
The similar idea has been applied to simple power grid simulation 
with matrix exponential method
\cite{Zhuang13_ASICON}.
Here, we generalize this technique and integrate into MATEX.
The Arnoldi process constructs $\mbf{V_m}$ and $\mbf{H_m}$, 
and the relationship is given by
$
    (\mbf{I}-  \gamma \mbf {A})^{-1} \mbf{V_m} =
    \mbf{V_m}\mbf{\widetilde{H}_m} + \tilde{h}_{m+1,m} \mbf v_{m+1} \mbf e^T_m, 
    $
we can project the $e^{\mbf{{A}}}$ onto the rational Krylov subspace as follows.
\begin{eqnarray}
    \label{eq:scale}
    e^{\mbf{A}h} \mbf{v}  \approx
    \left\| \mbf v \right\| \mbf{V_m} e^{h\mbf{  H}_m}e_1,
\end{eqnarray}
where 
${ \mbf{H_m}} = \frac{\mbf I - \mbf {\widetilde{H}}_m^{-1}}{\gamma}$ for the line \ref{alg1:h} of Alg. \ref{algo:arnoldi}. 
Following the same procedure \cite{Botchev12}, 
posterior error approximation is derived as  
\begin{eqnarray}
\label{eq:err_rational_krylov}
\lVert \mbf r_m(h) \rVert 
=
\lVert \mbf v \rVert
\left |  
\frac{\mbf I - \gamma \mbf A}{\gamma} 
\tilde{h}_{m+1,m}
\mbf v_{m+1} 
\mbf e^T_m \mbf{\widetilde {H}}_m^{-1}  
 e^{ h\mbf{ H}_m} 
 \mbf e_1 \right |
\end{eqnarray}
Note that in practice, instead of 
computing $(\mbf I-\gamma \mbf{A})^{-1}$ directly,
$(\mbf C + \gamma \mbf G)^{-1}\mbf C $ is utilized. 
The corresponding Arnoldi process shares  
the same skeleton of Alg. \ref{algo:arnoldi} and Alg. \ref{algo:ckt_solver} with input matrices 
$\mbf X_1 = (\mbf C+ \gamma \mbf G)$ for the LU decomposition,
and $\mbf X_2 = \mbf {C}$.

\subsubsection{Regularization-Free Matrix Exponential Method}
When dealing singular $\mbf C$, MEXP needs the regularization process
\cite{Chen12_TCAD} to remove the singularity of DAE in Eq. (\ref{eqn:dae}).
It is because MEXP is required to factorize $\mbf C$ in Alg. 1. 
This brings extra computational overhead when the case is large. 
Actually, it is not necessary if we can obtain the generalized eigenvalues 
and corresponding eigenvectors for matrix pencil $(-\mbf G, \mbf C)$. 
Based on \cite{Wilk:79}, we derive the following lemma,
\begin{lemma}
Considering a homogeneous system 
$\mbf C \mbf {  \dot{x}} = -\mbf G \mbf x$, 
$\mbf u$ and $\lambda$ are the eigenvector and eigenvalue 
of matrix pencil $(-\mbf G, \mbf C)$, then $\mbf x =  e^{t\lambda} \mbf u$ is a solution of this system.
\end{lemma}
An important observation is that 
we can remove such regularization process out of MATEX, 
because during Krylov subspace generation, 
there is no need of computing $\mbf C^{-1}$ explicitly.
Instead, we factorize $\mbf G$ for inverted Krylov subspace basis generation (I-MATEX), 
or $(\mbf C+\gamma\mbf G)$ for rational Krylov basis (R-MATEX).
Besides, $\mbf H'_m$ and $\mbf {\wtd H_m}$ are   invertible, 
which contain corresponding important generalized eigenvalues/eigenvectors from 
matrix pencil $(-\mbf G, \mbf C)$, and define the behavior of linear dynamic system
in Eq. (\ref{eqn:dae}). 

In term of error estimation, because  $\mbf C$ is singular, $\mbf A$ cannot be formed explicitly. 
However, for a certain lower bound of basis number, these two Krylov subspace methods
begin to converge, and 
the error of matrix exponential approximation is reduced quickly. 
Empirically, the estimation can be replaced with 
$\lVert r_m(h) \rVert = \lVert \mbf v \rVert \left|  h_{m+1,m} 
\mbf e^{\mathbf T}_m e^{h\mbf{H}_m} \mbf e_1 \right |$
to approximate Eq. (\ref{eq:err_inverted_krylov}), where
$\mbf H_m = \mbf {H'}^{-1}_m, h_{m+1,m} = h'_{m+1,m}$
for inverted Krylov method;
$\mbf H_m = \frac{\mbf I - \mbf {\wtd H}^{-1}_m}{\gamma}, h_{m+1,m} = \wtd{ h}_{m+1,m}$
for rational Krylov method.

\begin{figure}[h]
    \centering
    \includegraphics[ width=3.4in]{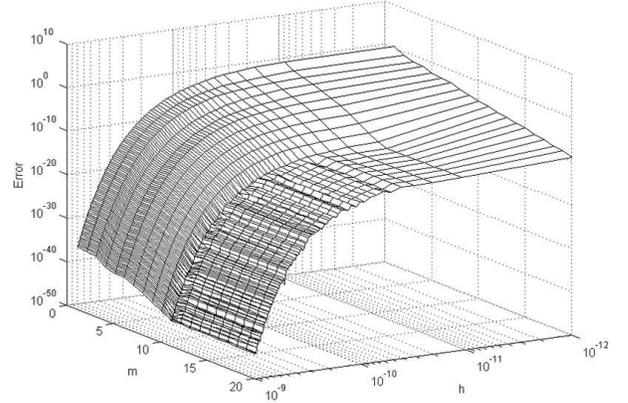}
    \caption{
	{\scriptsize
	$|e^{h\mbf A}\mbf v - \mbf V_m e^{h\mbf H_m}\mbf e_1|$
	    vs. time step $h$ and dimension of rational Krylov subspace basis (m). 
        $\mbf H_m$ = $\frac{\mbf I - \mbf {\wtd H}^{-1}_m}{\gamma} $; $\gamma $ is fixed; 
         $\mbf A$ is a relative small matrix and computed by MATLAB $expm$ function; 
	 Therefore, $e^{h\mbf A}\mbf v$ serves as the baseline for accuracy. 
	 It is observed that error reduces when $h$ increases.
	}
    }
    \label{fig:h_m}
\end{figure}

Note that the larger step R-MATEX utilizes, the smaller error it will have.
Fig. \ref{fig:h_m} shows, when time step $h$ increases, the error 
between accurate solution and Krylov based approximation 
$|e^{h\mbf A}\mbf v - \mbf V_m e^{h\mbf H_m}\mbf e_1|$ is reduced.
It is because the large step, the more dominating role first smallest magnitude eigenvalues play,
which are well captured by rational Krylov subspace-based method \cite{Van06}.
In our MATEX, this property is very crucial factor for large time stepping.
Therefore, once we obtain an accurate enough solution and 
Krylov subspace in line \ref{algo2:arnoldi} of Alg. \ref{algo:ckt_solver},
we can reuse them in line \ref{alg1:h}.



\subsection{Complexity Analysis}
\label{sec:complexity}

Suppose on average 
we have 
Krylov subspace basis dimension $m$ at each time step along the time span,
one pair of forward and backward   substitutions has time complexity $T_{bs}$. 
The matrix exponential evaluation using $\mbf H_m$ is $T_{H}$ which costs time complexity $O(m^3)$, plus extra $T_e$ to form $\mbf x$, which costs $O(n m^2)$.
The total time complexity of other serial parts is $T_{serial}$, which includes matrix factorizations, etc.
Given $K$ points of $GTS$, without decomposition of input transitions,
the time complexity is 
$KmT_{bs} + K (T_{H}+ T_e)  + T_{serial}$.
After dividing the input transitions and send to enough computing nodes, 
we have $k$ points of LTS for each node based on the input feature extraction and grouping 
(e.g., $k=5$ for one ``bump'' shape feature).
The total computation complexity is 
$kmT_{bs} + K (T_{H}+ T_e)  + T_{serial}$, where  $K (T_{H}+ T_e)$ contains the portion of computing for $Snapshot$.
The speedup of distributed computing over single MATEX is, 
\begin{eqnarray}
Speedup=\frac{ KmT_{bs} + K(T_{H}+ T_e)+ T_{serial}}  
{ kmT_{bs}+ K(T_{H}+ T_e)+T_{serial}}
\end{eqnarray}
In R-MATEX, we have very small $m$. Besides, $T_{bs}$ is larger than $T_{H}+T_e$. 
Therefore, the most dominating part is the $KmT_{bs}$.
We can always decompose input transitions, and make $k$ very small compared to $K$.
Traditional method with fixed step size has $N$ steps for the whole simulation.
The complexity is $NT_{bs}+T_{serial}$. Then the speedup of distributed MATEX over the one with fixed step size is,
\begin{eqnarray}
\label{eqn:tr_spd}
 Speedup'=\frac{NT_{bs}+T_{serial} }  { kmT_{bs}+ K(T_{H}+T_e)+T_{serial}}
\end{eqnarray}
Usually, $N$ is much larger than $K$ and $km$. 
Uniform step sizes make $N$ increased due to resolution of input transitions, 
to which $K$ is not so sensitive.
As mentioned before, $k$ can be maintained in a small number.
When elongating time span of simulation,
$N$
will increase.
However, $k$ will not change due to its irrelevant to time span 
(may bring more input transition features and increase computing nodes),
and then $Speedup'$ tends to become larger.
Therefore, our MATEX has more robust and promising theoretical speedups.



\section{Experimental Results}
\label{sec:exp_results} 

We implement our proposed MATEX in MATLAB R2013a and use UMFPACK package for LU factorization. 
The experiment is carried on Linux workstations with 
Intel Core\texttrademark~i7-4770 3.40GHz processor and 32GB memory on each machine.

\subsection{MATEX's Circuit Solver Performance} 
We test the part of circuit solver within MATEX 
using MEXP~\cite{Weng12_TCAD}, as well as our proposed I-MATEX and R-MATEX.
We create stiff RC mesh cases 
with different stiffness 
by changing the entries of matrix $\mbf C,~ \mbf G$. 
The stiffness here is defined as 
$ \frac{Re(\lambda_{min})}{Re(\lambda_{max})}$,
where $\lambda_{min}$ and $\lambda_{max}$ 
are the minimum and maximum eigenvalues of $-\mbf C^{-1} \mbf G$.
Transient results are simulated in $[0,0.3ns]$ with time step $5ps$.
Table~\ref{tab:rc_ckt} shows 
the average Krylov subspace basis dimension $m_{a}$ and the peak dimension $m_{p}$
used to compute matrix exponential during the transient simulations. 
We also compare the runtime speedups (Spdp) over MEXP. 
Err (\%) is the relative error opposed to 
BE method with a tiny step size $0.05ps$.
\begin{table}[ht]
    \caption{Comparisons among MEXP, I-MATEX and R-MATEX with RC cases}
    \centering
    \begin{tabular}{|c||c|c|c|c|c|c|}
    \hline
	Method  & $m_{a}$ &   $m_{p} $  & Err(\%)    & Spdp 
	& {Stiffness}  \\ 
	\hline
	MEXP	&  211.4   	&   229  	&  0.510   & -- &
	\multirow{3}{*}{$2.1 \times 10^{16}$ } 
	\\
	\cline{1-5}
	I-MATEX	& 5.7 	&14       &  0.004	   & 2616X   &   \\
	\cline{1-5}
	R-MATEX	& 6.9   & 12   &  0.004   &  2735X   & \\ 
	\hline
	\hline
	MEXP	&  154.2   	&   224  	&  0.004 & --  
	& 	    \multirow{3}{*}{$2.1 \times 10^{12}$ } 
	\\
	\cline{1-5}
	I-MATEX	&  5.7 	& 14       &0.004   &  583X  &   \\
	\cline{1-5}
	R-MATEX	& 6.9  & 12   &  0.004 	   &  611X  & \\ 
	\hline
	\hline
	MEXP &  148.6   	&   223  & 0.004  &  -- & 	    \multirow{3}{*}{$2.1 \times 10^{8}$ } 
	\\
	\cline{1-5}
	I-MATEX	& 5.7  	& 14      & 0.004     & 229X  &   \\
	\cline{1-5}
	R-MATEX	& 6.9  & 12   &  0.004  &  252X  & \\ 
	\hline
    \end{tabular}
    \label{tab:rc_ckt}
\end{table}
\begin{table*}[t]
 \centering
    \caption{TR with adaptive stepping (TR(adpt)) vs. I-MATEX vs. R-MATEX.
    ``Total(s)'' is the total runtime of test cases.
    ``DC(s)'' records the time to obtain initial condition.
    Spdp$^1$: Speedup of I-MATEX over TR(adpt); 
    Spdp$^2$: Speedup of R-MATEX over TR(adpt);  
    Spdp$^3$: Speedup of R-MATEX over I-MATEX.  
   }
    \begin{tabular}{|c|r|r||r|r|| r| r|  r| }
    \hline
    \multirow{2}{*}{Design} & 
    \multirow{2}{*}{DC(s)} &
    \multicolumn{1}{c||}{ TR(adpt)} & 
    \multicolumn{2}{c||}{I-MATEX} & 
    \multicolumn{3}{c|}{R-MATEX } 
    \\ 
    \cline{3-8} 
    &
			     &Total(s) 
			     &Total(s) 
			     &SPDP$^1$
			     &Total(s) 
			     &SPDP$^2$
			     &SPDP$^3$
			     \\ \hline  
	ibmpg1t		     & 0.13     
			      & 29.48
			      & 27.23
			     & 1.3X
			     & 4.93	  
			     & 6.0X 
			     & 5.5X
			     \\ \hline 
    	ibmpg2t 	     & 0.80  
			    & 179.74 
			   &  124.44
			   & 1.4X 
			   & 25.90  
			   & 6.9X 
			   & 4.8X
			    \\ \hline 
    	ibmpg3t 	     & 13.83 
			    & 2792.96 
			   & 1246.69  
			   & 2.2X
			   & 244.56  
			   & 11.4X
			   & 5.1X
			   \\ \hline
    	ibmpg4t 	    & 16.69
			    & 1773.14 
			    & 484.83 
			   & 3.7X 
			    & 140.42 
			    & 12.6X
			    & 3.5X
			    \\ \hline
    	ibmpg5t 	    & 8.16 
			   & 2359.11
			    & 1821.60 
			    & 1.3X 
			    & 337.74 
			    & 7.0X
			    & 5.4X 
			    \\ \hline 
    	ibmpg6t 	     & 11.17
		    	& 3184.59
			    & 2784.46 
			   & 1.1X
			    & 482.42  
			    &   6.6X 
			   &5.8X 
			    \\ 
    \hline
    \end{tabular}
    \label{tab:perf_mexp}
\end{table*}
\begin{table*}[tp]
\centering
\caption{MATEX vs. TR ($h=10ps$); 
Max. and Avg. Err.: maximum and average differences compared
to all output nodes' 
solutions provided by IBM Power Grid Benchmarks;
Spdp$^4$: ${t_{1000}/t_{rmatex}}$ transient stepping runtime speedups of MATEX over TR;
Spdp$^5$: $ {t_{t\_total}/t_{r\_total}}$ total simulation runtime speedups
of MATEX over TR. 
}
\begin{tabular}{|c||r|r||r|r| r| r|r|r|r|}
\hline
\multirow{2}{*}{Design}&
\multicolumn{2}{c||}{TR}&
\multicolumn{7}{c|}{MATEX}
\\
\cline{2-10}
& $t_{1000}(s)$
& $t_{t\_total}(s)$
& Group \#
& $t_{rmatex}(s)$ 
& $t_{r\_total}(s)$
& Max. Err.
& Avg. Err.
& Spdp$^4$
& Spdp$^5$
\\
\hline
 ibmpg1t & 5.94   &   6.20  &  {\color{black} 100}  &  0.50  &  0.85  &  1.4E-4 & 2.5E-5 & 11.9X  & 7.3X  
\\
\hline
 ibmpg2t & 26.98  &   28.61  &  100  &  2.02  &  3.72  &  1.9E-4 &  4.3E-5 & 13.4X  & 7.7X  
\\
\hline
 ibmpg3t & 245.92  & 272.47  &  100  &  20.15  &  45.77 &  2.0E-4 &  3.7E-5 & 12.2X  & 6.0X  
\\
\hline
 ibmpg4t & 329.36  & 368.55  &  {\color{black} 15} & 22.35 &  65.66 &  1.1E-4 &  3.9E-5 & 14.7X  & 5.6X  
\\
\hline
 ibmpg5t & 408.78  & 428.43  &  100 & 35.67 & 54.21 &  0.7E-4 &  1.1E-5 &  11.5X  &  7.9X  
\\
\hline
 ibmpg6t & 542.04  & 567.38  &  100 & 47.27 & 74.94 &  1.0E-4 &  3.4E-5 &  11.5X  & 7.6X  
\\
\hline
\end{tabular}
\label{tab:pg_rmatex}
\end{table*}

The huge speedups by I-MATEX and R-MATEX 
are due to large reductions of Krylov subspace basis $m_a$ and $m_p$. 
As we known, MEXP is good at handling mild stiff circuits \cite{Weng12_TCAD}, 
but inefficient on highly stiff circuits. 
Besides,
we also observe even the basis number is large, 
there is still possibility with relative larger error
compared to I-MATEX and R-MATEX. 
The average dimension $m_a$ of R-MATEX is a little bit larger than I-MATEX. 
However, the dimensions of R-MATEX used along time points spread more evenly than I-MATEX.
In such small dimension Krylov subspace computation, the total simulation runtime tends to be 
dominated by matrix exponential evaluations on the time points with the peak basis dimension($m_p$),
which I-MATEX has more than R-MATEX.
These result to a slightly better runtime performance of R-MATEX over I-MATEX.
In large scale of linear circuit system and practical VLSI designs, 
the stiffness may be even more extensive and complicated. 
Many of them may also have large singular $\mbf C$ 
and MEXP cannot handle without regularization process.
These make  I-MATEX and R-MATEX good candidates to deal with these scenarios.


\subsection{Adaptive Time Stepping Comparisons}

\vspace{-0.02in}
IBM power grid benchmarks~\cite{Nassif08_Power} are used to
investigate the performance of adaptive stepping TR (adpt) based on LTE controlling 
\cite{Weng12_TCAD, Najm10_book} as well as the performance of I-MATEX  and R-MATEX. 
Experiment is carried out on a single computing node.
In Table \ref{tab:perf_mexp}, the speedups of I-MATEX is not as large as R-MATEX because 
I-MATEX with a large spectrum of $\mbf A$ generates large dimension of Krylov subspace. 
In \emph{ibmpg4t} case, I-MATEX and R-MATEX achieve maximum speedups resulted from 
relative small number points in $GTS$, which around $44$ points, 
while the majority of  others  have over $140$ points.

\vspace{-0.02in}
\subsection{Distributed MATEX Performance}

\vspace{-0.02in}
We focus on MATEX with R-MATEX in the following experiments
with IBM power grid benchmarks. 
These cases have many input transitions ($GTS$) 
that limit step sizes of MATEX.
Exploiting distributed computing,
we decompose the input transitions, 
to obtain much fewer transitions of $LTS$ for computing nodes.
The input sources number is over ten thousand in the benchmarks, 
however, based on ``bump'' feature, we obtain a fairly small number
of the required computing nodes, which is shown as \emph{Group \#} in Table. \ref{tab:pg_rmatex}.
To compete the baseline classical TR method with fixed time step $h=10ps$, which requires 1000 pairs of  forward and backward substitutions
for the transient computing after factorizing $(\mbf C/h+\mbf G/2)$.
In R-MATEX, $\gamma =  10^{-10}$ is set to sit among the order of 
varied time steps during the simulation.
First, we pre-compute $GTS$ and $LTS$ groups and assign subtasks to corresponding nodes.
MATEX scheduler is only responsible for simple superposition calculation at the end of simulation.
Since MATEX slave nodes  are in charge of  
all the computing procedures (Fig. \ref{fig:matex}) for transient simulation,
and have no communications with each other during transient simulations,
we can easily emulate such multiple-node environment using our workstations.
We assign one MATLAB instance at each node of our workstations. 
After all MATEX slave nodes finish their jobs, we report the maximum runtime 
among these nodes as the total runtime $t_{r\_total}$ of MATEX.
We also record ``pure transient computing'', 
the runtime of transient part $t_{1000}$  and $t_{rmatex}$ excluding LU,  
where $t_{rmatex}$ is the maximum runtime of the counterparts among all  MATEX nodes.

Our MATEX framework achieves  13X on average
with respect to the pure transient computing $t_{1000}/t_{rmatex}$
as well as 7X on the total runtime $t_{t\_total}/t_{r\_total}$.
The average number of pairs of forward and  backward  substitutions
for Krylov subspace generations
is around 60 ($km$ in Eq. (\ref{eqn:tr_spd})),
while TR($h=10ps$) has 1000 pairs ($N$ in Eq. (\ref{eqn:tr_spd})) on each cases.
The reductions of these substitutions bring large speedups in the pure transient computing. 
With huge reductions on these substitutions, the serial parts, including the operations for LU and DC, play more dominating roles in 
MATEX, which can be further improved by other more advanced methods. 
\section{Conclusions}
\label{sec:conclusion}

We proposed a distributed framework MATEX for PDN transient simulation using the matrix exponential kernel.
MATEX leverages the linear system's superposition property,
and decomposes the task based on input sources features 
in order to reduce computational overheads for its subtasks at different nodes. 
We also address the stiffness problem for matrix exponential based circuit solver by rational Krylov subspace (R-MATEX), which has the best performance in this paper for adaptive time stepping without extra matrix factorizations.
In IBM power grid benchmark,
MATEX achieves 13X speedup over the fixed-step trapezoidal framework on average in transient computing after its matrix factorization. The overall speedup is around 7X.

\section{Acknowledge}
We acknowledge the support from NSF-CCF 1017864.
We also thank Ryan Coutts, Lu Zhang and Chia-Hung Liu for their helpful discussions.

{
{
\bibliographystyle{abbrv}
\bibliography{DAC_MATEX_final}  
}
}

\end{document}